\documentclass[conference]{IEEEtran}
\IEEEoverridecommandlockouts
\usepackage{cite}
\usepackage{amsmath,amssymb,amsfonts}
\usepackage{algorithmic}
\usepackage{graphicx} 
\usepackage{subfigure}
\usepackage{textcomp}
\def\BibTeX{{\rm B\kern-.05em{\sc i\kern-.025em b}\kern-.08em
    T\kern-.1667em\lower.7ex\hbox{E}\kern-.125emX}}
\begin{document}
\title{ Employ Multimodal Machine Learning for Content quality analysis}
\author{\IEEEauthorblockN{Eric Du}
	\IEEEauthorblockA{Key Laboratory of Trustworthy \\ Distributed Computing \\and Service (BUPT)\\Ministry of Education\\
		Beijing University of Posts \\  and Telecommunications\\
		Beijing, P.R.China\\
		Email: dupf@bupt.edu.cn}\and
	\IEEEauthorblockN{Xiaoyong Li}
	\IEEEauthorblockA{Key Laboratory of Trustworthy \\ Distributed Computing \\and Service (BUPT) Ministry of Education\\
		Beijing University of Posts \\ and Telecommunications\\
		Beijing, P.R.China\\
		Email: lxyxjtu@163.com}\and
 }
\maketitle

\graphicspath{{figures/}}
\begin{abstract}
The task of identifying high-quality content becomes increasingly important,
 and it can improve overall reading time and CTR(click-through rate estimates).
Generalizes quality analysis only focused on single Modal,such as image or text,but in today's 
mainstream media sites a lot of information is presented in graphic form.In this paper we propose 
 a MultiModal quality recognition approach for the quality score.
 First we use two feature extractors,one for image and another for the text, 
 After that we use an Siamese Network with the rank loss as the optimization objective.Compare with
 other approach,our approach get a more accuracy result.
\end{abstract}

\section{Introduction}
\par In Recent years there are greate change in the mainstream website,different with
the traditional published material,user-generated content has become increasely
popular on the web,more and more users participate in content creation,
With the coming of new media age,some Website such as twitter,weibo,toutiao,the content generated by user becomes
more important.
But the quality of user-generated content varies from excellent to abuse and spam,
so it becomes very important to filter high quality content.
And meanwhile the content of the website is not longer stuck in text,
it becomes varied with different modal,such as image、video,
so we should propose a method with multiple modalities quality analysis.
\par There are multiple modalities in the real world, we can hear sounds,
we can smell odors,and we see objects,and so on. 
In general terms,a modality refers to the way in which something happens or is experience. 
There are many problems in this area, such as image caption,multimodel tags,translation and we need to measure similarity between different modalities
and deal with possible long-range dependency and ambiguities.
it is difficult to construct a joint representation invariant across different modalities and the rapid 
increase in multimedia data transmission over the Internet necessitates the multi-modal summarization (MMS) from collections of text, image, audio and video.
MultiModal Machine Learning aims to build models that can process 
and relate information from multiple modalities.
\par In this paper we focus on the problem how to 
evaluate the quality of the content in an multimedia environment.
The main contributions are summarized as follows:
\begin{itemize}
	\item Different from traditional single modal quality analysis method,We use multimodal quality analysis approach,the feature in different modals is extract separately by the good performance feature extractors, One for InceptionV3 \cite{inceptionv3} another for transform-XL \cite{transformer-xl}.
	And we concat the two output representation as one output.
	\item We use the siamese network and rankloss for the final quality score,which the rank loss can learn the different
	between the high quality content and low quality content.
	\item Because the content may contains different number of pictures,so we use NeXtVLAD \cite{NeXtVLAD} 
	to aggregate frame-level features into a compact feature vector.
	\end{itemize}

\par The rest of this paper is organized as follows. 
Section 2 summarizes the related work. 
Section 3 introduces our architecture of multimodal content analysis model in details. 
Section 4 presents the experimental results. Finally we conclude in Section 5

\section{Related Work}
\subsection{Multimodal Machine Learning }
The multimodal machine learning aims to build models that can process information from multiple modalities. 
multimodal machine learning is a vibrant multi-disciplinary field of increasing importance and with extraordinary potential.
In recent year there are there category of multimodal application 
One category is multimedia content indexing and retrieval, In the paper \cite{mms}, 
it produced summaries, based on low-level features and content-independent fusion and selection, 
are of subjectively high aesthetic and informative quality.There are also a multimedia event detection (MED) tasks started in 2011 \cite{RECVID}.
The second category is multimodal interaction with the goal of understanding human
multimodal behavior,such as emotion recognition,there are many dataset Corpus,one is 
SEMAINE corpus which allowed to
 study interpersonal dynamics between speakers and listeners \cite{emotiondataset}
 This dataset formed the basis of the first audio-visual emotion challenge (AVEC) 
 organized in 2011 \cite{avec}.
 The AVEC challenge with healthcare applications such as automatic assessment of depression and anxiety \cite{atext}.
 Another  category of multimodal application  something about the video ,because video is an multimedia ,it contains text,image,and voice.
 There are dataset for video classification,such as YouTube-8M: A Large-Scale Video Classification Benchmark\cite{video-y8m}
Another category of multimodal applications with an emphasis on language and vision is
media description,the most representative application is image caption
including the multimedia event detection (MED) tasks started in 2011 
A new category of multimodal is media description.It emerged with an emphasis on language and vision.
The most representation task is image caption,it generate a text description of input image \cite{imagecaption}.
another representation task is visual question answering.It was recently proposed to 
address some of the evaluation challenges \cite{vqa}.

\subsection{Content Quality analysis}
Most scholars focus on image quality or text quality,
Le Kang et al. \cite{CNN-NRIQA} 
used 5-layer CNN to accurately predict NR-IQA, 
This method performs best on LIVE datasets at the time,And the experiment of local image distortion proves the local quality estimation ability of CNN.
Xialei Liu et al. \cite{RankIQA} proposed RankIQA model to evaluate the quality of images without reference.
Firstly, the author selects the representation characteristics of the data ranking relations generated by Siamese network learning, and then transfers the knowledge expressed in the trained Siamese network to the traditional CNN, so as to estimate the absolute image quality of a single image. 
In the field text quality assessment, Naderi et al. \cite{Automated-text-readability-assessment}
has developed an supervised learning algorithms over german text corpora, it extract
73 linguistic features grouped in traditional, lexical and morphological features.
The results obtained depict that Random Forest Regressor yielding best result (0.847) for RMSE measure.
Mesgar et al.\cite{Neural Local Coherence Model} proposed  a local coherence model that cap-tures theflow of what semantically connectsadjacent sentences in a text. 

\section{Method}
\subsection{Overview of our approach}
\begin{figure}[htbp]
	\centering
\includegraphics[scale=.2]{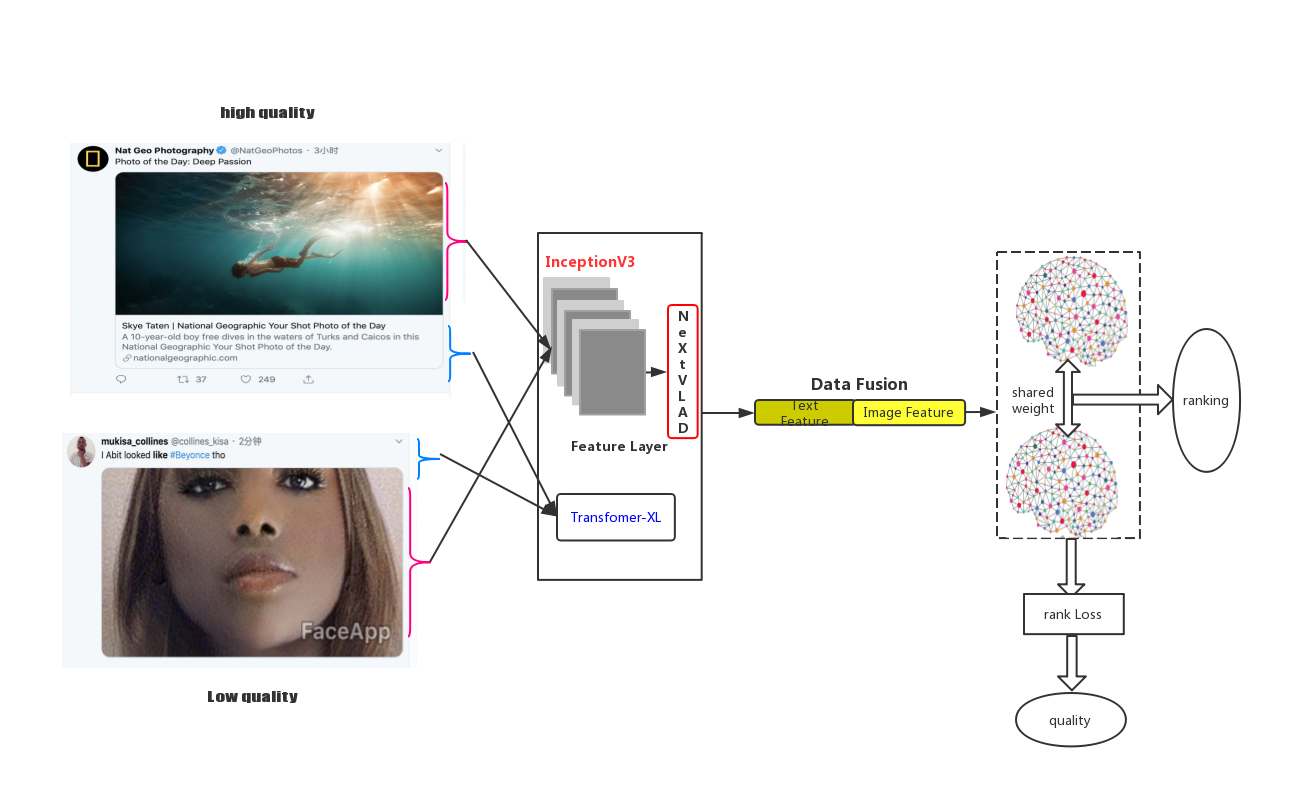}
\caption{In this approach we first collect a graphic set of high quality and low quality,we use 
	the Siamese networks which is used to learning rank from different quality.	}
\label{fig:1} 
\end{figure}

The content quality in this paper mainly considers the content of blogs combined with pictures and texts, and this kind of content is the main output mode .
The traditional way of processing is to consider from the perspective of both pictures and text. The quality of traditional pictures is mainly evaluated by the blur, blockiness, and the severity of the noise, and evaluate the degree of distortion. 
The main underlying features are color, hue, contrast, corner, texture, etc.
Combine some high-level features such as aesthetics, attraction, etc., and then get a final image quality score through a deep neural network.
For text quality, it is mainly based on textuality, length, abnormal character proportion, type label, part of speech, dependency, syntactic structure and other characteristics.
In this paper we proposed an approach to solve the disadvantages of information missing in single dimension. 
The main idea is adopts a unified fusion architecture to fuse text features and image features (see \figurename{fig:1} ). 
We use transform-XL for the text feature extract, it can solve the limitation of the fixed length context.
and use InceptionV3 for the image feature extract,and then we concat the text feature and the image feature.Then the full-connected layer as the rank layer. 
\subsection{Text feature extract with transformer-XL}
In order to learn more hidden information especially for more context information,we use transform-XL \cite{transformer-xl}.
Compared with the Transformer,Transformers-XL can conquer the limitation of fixed-length context in the
setting of language modeling.In Transformers-XL it reuse the hidden states obtained in previous segments. 
For the current segment the reused hidden states serve as memory.Transformers-XL using relative positional encoding
formulation that generalizes to attention lengths.
\subsection{The Siamese network for content quality ranking}
\par We use the siamese network to learn from content ranking,
the network has two identical network brach and a loss module. 
And the two branches share weights during training. 
We use pairs of content and labels as the input of the network. 
The loss function is the ranking loss which we can learning the difference quality,we compute the
gradients of the loss function with the Adaptive Moment Estimation method ,
In essence, RMSprop with momentum items uses the first-order moment estimation of the gradient and the second-order moment estimation to dynamically adjust the learning rate of each parameter. 
Adam's advantage is that after bias correction,
making the parameters relatively smooth,
the learning rate for each iteration has a definite range.
 \par Given an content ( text with image ) as the input of the network, the output feature representation of X, 
 is obtained by combination of image representation and text representation. 
 Assume that $X_{A}$ for high quality content
 and $X_{B}$ for the low quality content. The scoring model is used to evaluate the probability of the quality
  that $X_{A}$ is higher than $X_{B}$ . We calculate the probability of ranking in pair.Formula is as follows:
  \begin{equation}
	P_{i j} \equiv P\left(U_{i} \oslash U_{j}\right) \equiv \frac{1}{1+e^{-\sigma\left(X_{i}-X_{j}\right)}}
  \end{equation}
The ground truth probability as follows:
\begin{equation}
	\overline{P}_{i j}=\frac{1}{2}\left(1+X_{i j}\right)
\end{equation}
In order to measure the fitting degree of Pij and $\overline{P}_{i j}$ we should use the loss function,
Because of the normalization operation, the traditional mean-square loss function has become non-convex，
so we use the cross entropy loss,When the two content are in different quality, 
it gives some punishment to let them separate.The Formula is as follow:
\begin{equation}
C_{i j}=-\overline{P}_{i j} o_{i j}+\log \left(1+e^{o_{i j}}\right)
\end{equation}
And the form of cross entropy is more suitable for gradient descent method.
\subsection{A method of graphic fusion}
\par
\begin{figure}[htbp]
	\centering
\includegraphics[scale=.25]{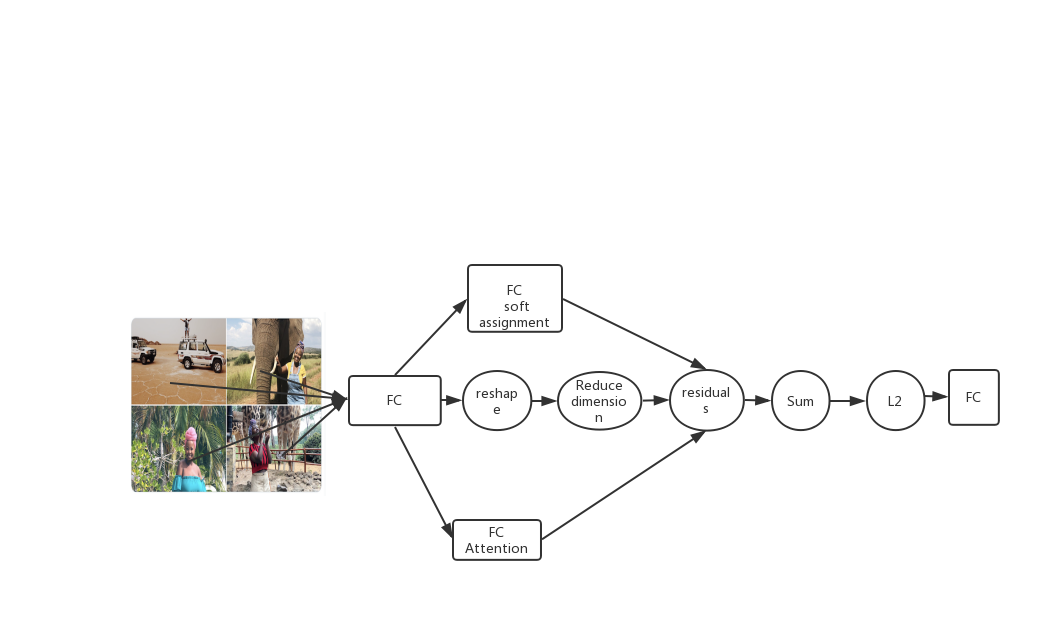}
\caption{NeXtVLAD for the original images input.}\label{fig:2} 
\end{figure}

 In order to identical the quality of the graphic content
 there are many problems need to solve,First of all,
 usually there are many pictures in each blog content,
 we need to fuse them into one embedding feature, 
 We use NeXtVLAD \cite{NeXtVLAD} to aggregate frame-level features into a compact feature vector.
 The main idea is to decompose a high-dimensional record
 vector into a group of low-dimensional vectors with attention before applying NetVLAD\cite{NetVLAD},
The operation of the NeXtVLAD is described as \figurename{fig:2}

The input images are output as fc layer,then a reshape operation is applied to transformer the organized images,
the processes is equivalent to spliting the output into lower-dimensional feature ,
each of which is subsequently represented as a mixture of residuals from cluster anchor point in the same low-dimensional
space:
 \begin{equation}
	p_{i j k}^{g}=\beta_{g}\left(\dot{x}_{i}\right) \beta_{g k}\left(\dot{x}_{i}\right)\left(\tilde{x}_{i j}^{g}-c_{k j}\right)
\end{equation}
 The first part $\beta_{g k}\left(\dot{x}_{i}\right)$ measures the soft assignment of $\tilde{x}_{i j}^{g}$ to the cluster k ,and the second
 part $\beta_{g}\left(\dot{x}_{i}\right)$ can be regarded as an attention function over all the pictures.
\par Then, the descriptor of many images is achieved via aggregating the encoded vectors over the images:
 \begin{equation}
 y_{j k}=\sum_{i, g} v_{i j k}^{g}
 \end{equation}
\par There are many method to contact from individual modalities into a joint space,
the previous surveys emphasizing early,late and hybrid fusion approach.
such as contact fusion,late fusion and early fusion,
and finally we fusion the different modalities with concat operation.


\section{Experimental AND RESULTS}
\subsection{Dataset}
\par
\begin{figure}[htbp]
	\centering    
	\subfigure[twitter content ] 
	{
		\begin{minipage}{7cm}
		\centering          
		\includegraphics[scale=0.35]{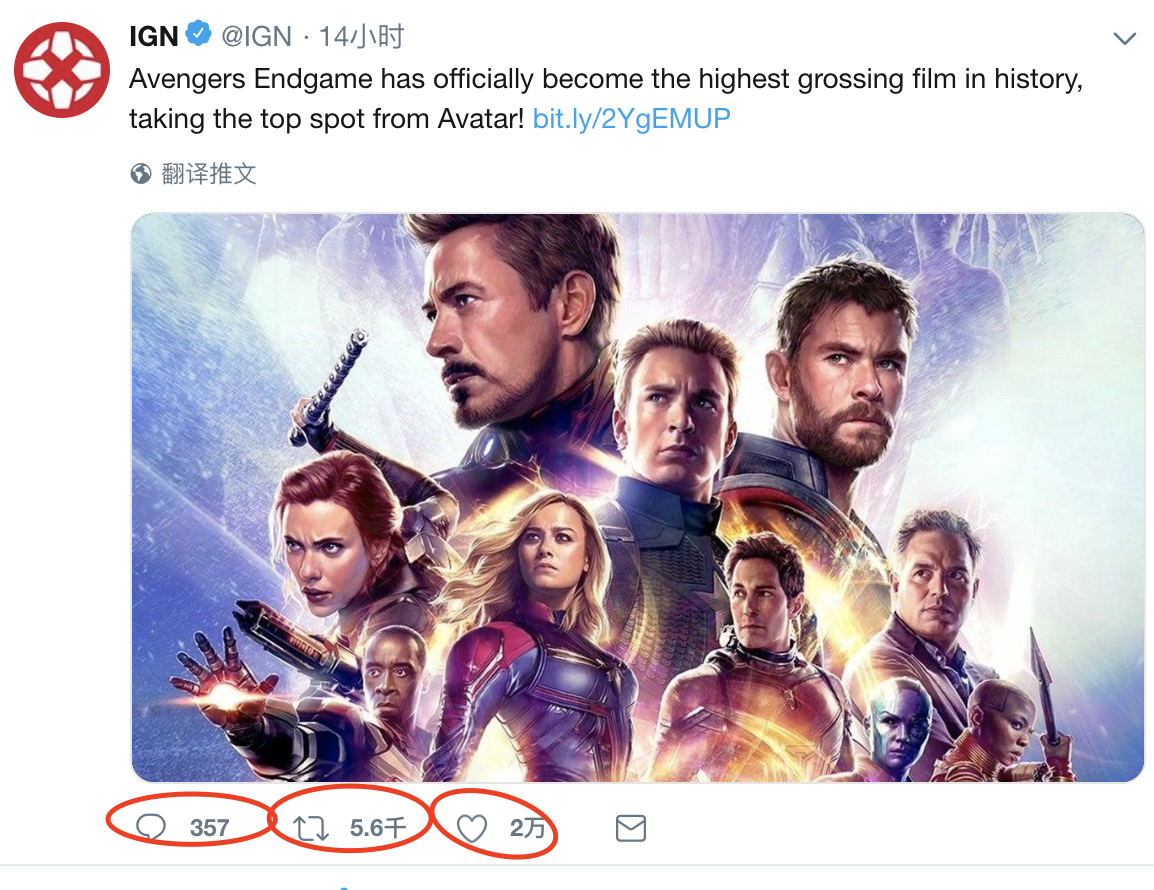}   
		\end{minipage}
	}
	\subfigure[weibo content ] 
	{
		\begin{minipage}{7cm}
		\centering      
		\includegraphics[scale=0.35]{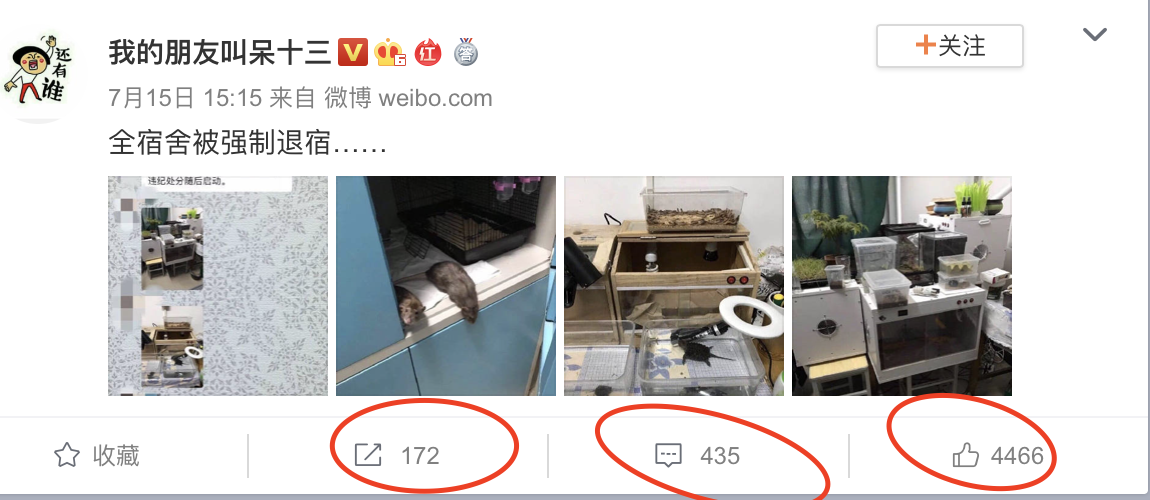}   
		\end{minipage}
	}
	\caption{retweets,comment,likes Feature for evalute the quality of the content} 
	\label{fig:4}  
	\end{figure}

 For evaluating our model, we crawled many data from different website,such as weibo、twitter, 
and most of them are presented in graphic form,In order to score the quality of the content ,we 
use the retweets,comment,likes feature .Supposed that if there are more 
Interactive behavior the quality of the tweet is high,it is show in \figurename{fig:4} 
and we also employ someone to 
evaluate the score in general,and weed out cases of obvious error.

	\begin{equation}
		LCC=\frac{\sum_{i=1}^{N}\left(y_{i}-\overline{y}\right)\left(\hat{y}_{i}-\hat{y}\right)}{\sqrt{\sum_{i}^{N}\left(y_{i}-\overline{y}\right)^{2}} \sqrt{\sum_{i}^{N}\left(\hat{y}_{i}-\overline{\hat{y}}\right)^{2}}}
	\end{equation}
Because the important of the retweets,comment,likes features are different, And  
more likes feature indicated more information. so we give different weights to different features.
\par
\subsection{Multimodal Quality Assessment}
\par We use Stochastic Gradient Descent 
with an learning rate of 1e-5 for efficient Siamese network training.
Traning rates are decreased by 
a factor of 0.08 every 10,000 iterations for a total of 50,000 iterations
As there are many images for each blog,we take 4 images for each blogs.
The evaluation metrics are traditionally used to evaluate the performance
of the algorithms is that the Linear Correlation Coefficient (LCC),
LCC is a measure of the linear correlation between the ground truth and the predicted quality scores. Given N Graphic, the ground truth of i-th image is denoted by yi, and the predicted score from the network is $^{\wedge} y_{i}$. 
And the formulation of LCC like that :
\begin{equation}
	LCC=\frac{\sum_{i=1}^{N}\left(y_{i}-\overline{y}\right)\left(\hat{y}_{i}-\overline{\hat{y}}\right)}{\sqrt{\sum_{i}^{N}\left(y_{i}-\overline{y}\right)^{2}} \sqrt{\sum_{i}^{N}\left(\hat{y}_{i}-\overline{\hat{y}}\right)^{2}}}
\end{equation}
\subsection{Comparision with other approach }
 We have made use of various classifiers such as Logistic Regression(LR),
Linear SVR (L-SVR),XGBOOST. 
They report the LCC values obtained by the classifier for different combination of
modalities.We also use deeplearning approach for the baseline,
With the method proposed by Xialei Liu et al.\cite{RankIQA} in NR-IQA, we use the approach for the 
single picture score.
Another approach for single modal recognition \cite{Transparent text} 
proposed a deep rank approach for text quality scores.
Another approach is CNN for image feature extraction,
and Bi-LSTM for text feature extraction,concat the feature with early fusion and the loss is Square Loss.
And we use the first picture for the feature extraction.

Table \ref{tab:table1}  \ref{tab:table2} shows that our model CNN and transform-XL with loss of rankloss,obtains a LCC of 0.913,
while outperforming all other competitive baselines.


\section{Conclusions And Future Work }
\par This paper addresses an Multimodal quality score task, 
namely, how to use related text,image to score the quality of content. 
First we extract the feature with cnn for image representation,and we extract the
feature with transform-XL for text representation,with the concat of the features,We use
rankloss for the final optimization objective,Experiments show that it is better than
the single modal recognition and the traditional deeplearning fusion approach.
\par In mainstream media website, video a growing proportion of the share, and in the
future work we should fusion the video features for the quality analysis.

\begin{table}
	\caption{Comparision of LCC Baseline}
	\label{tab:table1}
		\begin{tabular}{|l|c|c|c|}
			\hline
			\bf Features & LR& L-SVR & XGBOOST  \\
			\hline
			\bf Image(Single) & 0.6694 & 0.711 & 0.78  \\
			\bf Image(Multi) & 0.5661 & 0.6425 & 0.711   \\
			\bf Textual & 0.6457 & 0.5643 & 0.8231 \\
			\hline \hline
			\bf Multimodal & 0.501 & 0.579 & 0.611  \\
			\hline
	\end{tabular}

\end{table}

\begin{table}
	\caption{Comparision of LCC }
	\label{tab:table2}
		\begin{tabular}{|l||c|c|c|c|}
			\hline
			\bf Features & RankIQA\cite{RankIQA} & CNNtext\cite{Transparent text} & DL+SquareLoss & Our approach \\
			\hline
			\bf Image(Single)  & 0.80 & - & - & - \\
			\bf Image(Multi)  & 0.813  & - & -& -  \\
			\bf Textual  & - & 0.77 & - & - \\
			\hline \hline
			\bf Multimodal & - & - & 0.8879 & \bf0.913 \\
			\hline
	\end{tabular}

\end{table}

%


\begin{thebibliography}{1}
\bibitem{mms}
	G. Evangelopoulos et al., "Multimodal Saliency and Fusion for Movie Summarization Based on Aural, Visual, and Textual Attention," in IEEE Transactions on Multimedia, vol. 15, no. 7, pp. 1553-1568, Nov. 2013.
	doi: 10.1109/TMM.2013.2267205
\bibitem{RECVID}
	"TRECVID Multimedia Event Detection 2011 Evaluation, https://www.nist.gov/multimodal-information-group/" trecvid-multimedia-event-detection-2011-evaluation, accessed: 2017-01-21.
\bibitem{emotiondataset}
	G. McKeown, M. F. Valstar, R. Cowie, and M. Pantic, “The SE- MAINE corpus of emotionally coloured character interactions,” in IEEE International Conference on Multimedia and Expo, 2010
\bibitem{inceptionv3}
Xiaoling Xia, Cui Xu and Bing Nan, "Inception-v3 for flower classification," 2017 2nd International Conference on Image, Vision and Computing (ICIVC), Chengdu, 2017, pp. 783-787.

	\bibitem{avec}
	B. Schuller, M. F. Valstar, F. Eyben, G. McKeown, R. Cowie, and M. Pantic, “AVEC 2011 – The First International Audio / Visual Emotion Challenge,” in ACII, 2011
\bibitem{atext}
	M. Valstar, B. Schuller, K. Smith, F. Eyben, B. Jiang, S. Bilakhia, S. Schnieder, R. Cowie, and M. Pantic, “AVEC 2013 – The Contin- uous Audio / Visual Emotion and Depression Recognition Chal- lenge,” in ACM International Workshop on Audio/Visual Emotion Challenge, 2013.
\bibitem{video-y8m}
	Abu-el-haija, S., Lee, J., Natsev, P.,  Toderici, G. (n.d.). YouTube-8M : A Large-Scale Video Classification Benchmark.
\bibitem{imagecaption}
	M. Hodosh, P. Young, and J. Hockenmaier, “Framing image de- scription as a ranking task: Data, models and evaluation metrics,” JAIR, 2013
\bibitem{vqa}
	S. Antol, A. Agrawal, J. Lu, M. Mitchell, D. Batra, C. Lawrence Zitnick, and D. Parikh, “VQA: Visual question answering,” in ICCV, 2015.
\bibitem{transformer-xl}
	Dai, Z., Yang, Z., Yang, Y., Carbonell, J., Le, Q. V.,  Salakhutdinov, R. (2019). Transformer-XL: Attentive Language Models Beyond a Fixed-Length Context.
\bibitem{NetVLAD}
	Arandjelovic R, Gronat P, Torii A, et al. NetVLAD: CNN Architecture for Weakly Supervised Place Recognition[J]. IEEE Transactions on Pattern Analysis and Machine Intelligence, 2018, 40(6): 1437-1451.
\bibitem{NeXtVLAD}
	Lin, R., Xiao, J.,  Fan, J. (2019). NeXtVLAD: An efficient neural network to aggregate frame-level features for large-scale video classification. Lecture Notes in Computer Science (Including Subseries Lecture Notes in Artificial Intelligence and Lecture Notes in Bioinformatics), 11132 LNCS, 206–218. $https://doi.org/10.1007/978-3-030-11018-5_19$
\bibitem{RankIQA}
	Liu, X., \& Bagdanov, A. D. (2012). RankIQA : Learning from Rankings for No-reference Image Quality Assessment Supplementary Material.
\bibitem{Transparent text}
	Östling, R., \& Grigonyte, G. (2018). Transparent text quality assessment with convolutional neural networks, 282–286.
\bibitem{CNN-NRIQA}
L. Kang, P. Ye, Y. Li, and D. Doermann. Convolutional neural networks for no-reference image quality assessment. In Proceedings of the IEEE conference on computer vision and pattern recognition, pages 1733–1740, 2014.
\bibitem{Automated-text-readability-assessment}
Naderi, Babak, Mohtaj, Salar, Karan, Karan,\& Moller, Sebastian (2019). Automated Text Readability Assessment for German Language: A Quality of Experience Approach. 2019 Eleventh International Conference on Quality of Multimedia Experience (QoMEX), IEEE, doi:10.1109/qomex.2019.8743194

\bibitem{Neural Local Coherence Model}
Mesgar, Mohsen, \& Strube, Michael (2018). A Neural Local Coherence Model for Text Quality Assessment. Proceedings of the 2018 Conference on Empirical Methods in Natural Language Processing, Association for Computational Linguistics, doi:10.18653/v1/d18-1464

\end{thebibliography}
\end{document}